# Magnetic Particle Spectroscopy for Detecting Cell-Associated Zinc Ferrite Nanoparticles and Probing Their Relaxation Dynamics

Hanlei Wang[1,2,†], Bahareh Rezaei[3,†], Md Shahriar[4], Changxue Xu[4], Rui He[5], Kai Wu[1,*]

[1]Department of Physics, University of South Florida, Tampa, FL, USA

[2]Department of Electrical and Computer Engineering, University of South Florida, Tampa, FL, USA

[3]Department of Electrical and Computer Engineering, Texas Tech University, Lubbock, TX, USA

[4]Department of Industrial, Manufacturing, and Systems Engineering, Texas Tech University, Lubbock, TX, USA

[5]Department of Physics, The University of Texas at Arlington, Arlington, TX, USA

[†]Equal contribution

*Corresponding author E-mail: kaiwu@usf.edu (K.W.)

**ABSTRACT**

Magnetic particle spectroscopy (MPS) enables sensitive detection of magnetic nanoparticles (MNPs) and characterization of their dynamic magnetization through higher-order harmonics. Here, we synthesized citrate-functionalized, 30 nm cubic $Zn_{0.4}Fe_{2.6}O_4$ (ZFO) MNPs and investigated their association with SKOV3 ovarian cancer cells using MPS. The ZFO MNPs exhibited a crystalline spinel structure, a mean hydrodynamic diameter of 39.4 nm, strong room-temperature magnetization, low coercivity, and citrate-associated surface functional groups. Live/Dead imaging indicated good short-term cytocompatibility after 24 h exposure at concentrations up to 500 µg/mL, while bright-field microscopy showed concentration-dependent cell-associated nanoparticle accumulation. Two MPS drive fields were compared using ZFO MNPs dispersed in deionized (DI) water and immobilized in agar as relatively unrestricted and strongly confined reference states. The 7.75 kHz, 20 mT condition retained more higher-order harmonics than 11.37 kHz and 10 mT, as expected from its larger field amplitude, and, more importantly, produced a larger and order-dependent spectral separation between the water and agar states; it was therefore selected for the cellular measurements. After ZFO exposure and removal of unbound nanoparticles, MPS detected cell-associated nanoparticles in samples containing $0.1\times10^6$, $1\times10^6$, and $2\times10^6$ SKOV3 cells. The third-, fifth-, and seventh-harmonic amplitudes increased proportionally with cell number, with power-law exponents of 1.03-1.09. We also report that most normalized harmonic ratios from the $1\times10^6$- and $2\times10^6$-cell samples fell between the water and agar reference profiles, qualitatively indicating partial restriction of Brownian rotation in the cellular environment. These findings demonstrate that MPS can detect cell-associated ZFO MNPs while providing complementary spectral information about their ensemble-averaged physical confinement, supporting its application in magnetic cell labeling and cell-tracking studies.

## 1. INTRODUCTION

Magnetic particle spectroscopy (MPS) is a sensitive technique for characterizing the nonlinear dynamic magnetization of magnetic nanoparticles (MNPs). Following its first report in 2006[1,2] and subsequent establishment as a magnetic particle imaging (MPI) tracer-characterization method[3,4], MPS has evolved into a versatile platform for MNP analysis, biomedical sensing, and process monitoring. In MPS, an alternating magnetic field drives the MNP magnetization through the nonlinear region of the magnetization curve. The resulting time-dependent magnetization induces a voltage containing higher-order harmonics of the excitation frequency. Because diamagnetic and paramagnetic biological materials generate negligible higher-order nonlinear magnetic signals under typical MPS excitation conditions, these harmonics provide highly specific detection of MNPs with minimal biological background.[5,6] Within the linear operating range of the measurement system, the absolute harmonic amplitudes scale with MNP quantity, enabling quantitative detection and comparison of magnetic tracers.[7–9] Furthermore, the harmonic spectrum is also influenced by the magnetic and physicochemical properties of the particles, including core size, size distribution, saturation magnetization, magnetic anisotropy, hydrodynamic size, aggregation, and interparticle interactions.[10–12] Environmental parameters, including temperature[13–15], viscosity[16–19], and physical binding or immobilization[20–24], further alter Brownian and Néel relaxation and thereby change the amplitudes, phases, and relative strengths of the higher-order harmonics. These sensitivities have enabled MPS applications beyond particle quantification, including MPI tracer-performance screening, temperature and viscosity sensing, characterization of nanoparticle binding and aggregation[23–25], evaluation of blood-clot progression[26,27], and real-time monitoring of MNP synthesis[28,29].

MPS is useful for characterizing MNPs associated with biological cells because cellular interactions can alter both the amount and dynamic magnetic behaviors of the particles. After cellular association or internalization, MNPs may bind to the cell membrane, aggregate, or become confined within endosomes and lysosomes. These processes restrict whole-particle rotation and alter the relative contributions of Brownian and Néel relaxation compared with freely dispersed particles.[30–32] Löwa *et al*.[33] first demonstrated the feasibility of quantifying cellular MNP uptake using particle-specific MPS calibration curves, while Fidler *et al*.[34] applied MPS to assess the vitality of magnetically labeled stem cells. Poller *et al*.[30] subsequently showed that cellular interaction and internalization altered the $A_5/A_3$ harmonic ratio of citrate-coated very small superparamagnetic iron oxide nanoparticles in THP-1 cells. Incorporating this spectral change improved the accuracy of cellular MNP quantification. MPS has also been used to quantify superparamagnetic iron oxide nanoparticles (SPIONs) associated with and transported across human brain microvascular endothelial cell layers[9] and to characterize changes in MNP signal amplitude and spectral shape in different biological environments[31]. Subsequent studies confirmed that cellular processing, particle aggregation, cell type, and excitation-field conditions can affect both MPS amplitude and spectral shape.[35–38]

These previous findings reveal an important analytical challenge: the absolute MPS amplitude of a cell sample may be affected by both the amount of cell-associated MNPs and environment-induced changes in their relaxation dynamics. Therefore, a change in harmonic amplitude cannot necessarily be attributed to MNP amount alone. Normalized harmonic ratios provide complementary information because they describe spectral shape and, when the relevant harmonics remain above the noise floor, are comparatively insensitive to MNP quantity.[21,39–41] Harmonic ratios have been frequently used as MNP-amount-independent parameters that are only affected by the relaxation dynamics of MNPs, including the bound status of MNPs to analytes, the viscosity and temperature of the surrounding environment, etc.[18,25,31,42,43] Combining amount-sensitive harmonic amplitudes with relaxation-sensitive harmonic ratios may therefore help distinguish differences in total MNP content from differences in ensemble-averaged particle confinement. However, harmonic ratios must be interpreted cautiously at low MNP amounts because weak higher-order components can approach the noise floor and bias the calculated values. A reference-based approach that compares cellular harmonic ratios with relatively unrestricted and strongly immobilized particle states remains insufficiently explored.

In this study, we developed a reference-based MPS framework to evaluate both the amount-dependent and relaxation-dependent spectral characteristics of 30 nm cubic $Zn_{0.4}Fe_{2.6}O_4$ (ZFO) MNPs associated with SKOV3 ovarian cancer cells. Two drive-field conditions were first compared using ZFO MNPs dispersed in deionized (DI) water and immobilized in agar, and the condition providing stronger nonlinear excitation and greater higher-order-harmonic retention was selected for cellular measurements. Absolute third-, fifth-, and seventh-harmonic amplitudes were evaluated as functions of cell number, whereas several normalized harmonic ratios were used to compare spectral shape across samples containing different amounts of cell-associated ZFO. The harmonic ratios were subsequently compared between MNP-cell-associated samples and MNPs in water and agar reference profiles to qualitatively assess the ensemble-averaged degree of particle confinement. This study demonstrates a framework for jointly evaluating amount-dependent MPS signals and environmentally induced spectral changes in magnetically labeled cell samples.

## 2. MATERIALS AND METHODS

### 2.1. Materials

Ferric chloride hexahydrate ($FeCl_3 \cdot 6H_2O$, ≥98%), ferrous chloride tetrahydrate ($FeCl_2 \cdot 4H_2O$, ≥98%), zinc nitrate hexahydrate [$Zn(NO_3)_2 \cdot 6H_2O$, ≥98%], ammonium hydroxide ($NH_4OH$), citric acid (CA, ≥99.5%), and 200-proof ethanol were obtained from Sigma-Aldrich (St. Louis, MO, USA) or Fisher Scientific (Waltham, MA, USA). Phosphate-buffered saline (PBS), 0.25% trypsin solution, Dulbecco's modified Eagle medium (DMEM), fetal bovine serum (FBS), and antibiotics were used for cell culture. SKOV3 human ovarian cancer cells were obtained from the American Type Culture Collection (ATCC; Manassas, VA, USA). Cell viability was assessed using a Calcein-AM/EthD-1 Live/Dead assay kit (Biotium, Fremont, CA, USA). DI water was used to prepare all aqueous solutions. Unless otherwise specified, the reagents were used as received.

**2.2. Synthesis of ZFO MNPs**

Citrate-stabilized ZFO MNPs with a 30 nm cubic morphology were synthesized using a hydrothermal method. Briefly, $FeCl_3 \cdot 6H_2O$, and $Zn(NO_3)_2 \cdot 6H_2O$ were dissolved in DI water at molar proportions corresponding to the nominal $Zn_{0.4}Fe_{2.6}O_4$ composition. $NH_4OH$ was added to establish alkaline conditions and induce precipitation of the metal ions. After brief stirring at room temperature, the precipitate was isolated by centrifugation and repeatedly rinsed with DI water. The recovered solid was resuspended in DI water and mixed with citric acid for 30 min to promote citrate adsorption onto the particle surfaces. The suspension was subsequently sealed in a Teflon-lined stainless-steel autoclave and subjected to hydrothermal treatment at 195 °C for 16 h. After natural cooling, the nanoparticles were magnetically separated, washed with DI water and acetone, and vacuum-dried at 60 °C. The resulting powder was stored at room temperature. Before solution-based experiments, the nanoparticles were dispersed in DI water or the appropriate experimental medium by vortex mixing and sonication.

**2.3. Physicochemical Characterization of ZFO MNPs**

Particle morphology was characterized by transmission electron microscopy (TEM; Hitachi 7650). Dilute ZFO MNP suspensions were deposited onto carbon-coated copper grids and dried before imaging. The TEM images were used to evaluate particle size, shape, and aggregation. Crystal structure was analyzed by powder X-ray diffraction (XRD) using a Rigaku MiniFlex 6G diffractometer with Cu Kα radiation ($\lambda$ = 1.5406 Å). Dried nanoparticle powder was scanned over a 2θ range of 20-80° with a step size of 0.1° and a slit width of 1.25°. The measured diffraction pattern was compared with reference patterns to verify the formation of the spinel ferrite phase. Hydrodynamic diameter and polydispersity index were measured by dynamic light scattering (DLS; Microtrac Zetatrac NPA152). The nanoparticles were dispersed in DI water at 1 mg/mL and measured at room temperature after 30 minutes of sonication. Static magnetic properties were characterized at room temperature using a Physical Property Measurement System (PPMS; Quantum Design). Magnetization was recorded as a function of applied field up to 1.5 T. Saturation magnetization ($M_s$), coercivity ($H_c$), and remanent magnetization ($M_r$) were obtained from the resulting M-H loop. Fourier-transform infrared (FTIR) spectroscopy was used to examine the nanoparticle surface chemistry and confirm citrate association. Spectra of the dried nanoparticles were collected over 400-4000 $cm^{-1}$. Absorption bands associated with ferrite metal-oxygen vibrations and citrate functional groups were identified.

**2.4. Cytocompatibility and Cell-Associated Nanoparticle Accumulation**

The cytocompatibility of the ZFO MNPs was evaluated using SKOV3 ovarian cancer cells. Cells were maintained in DMEM supplemented with FBS and antibiotics at 37 °C in a humidified atmosphere containing 5% $CO_2$. After attachment, the cells were exposed to ZFO MNPs at final concentrations of 100 or 500 μg/mL for 24 h. Following exposure, the nanoparticle-containing medium was removed, and the cells were washed with PBS. Cell viability was qualitatively evaluated using the Calcein-AM/EthD-1 Live/Dead assay according to the manufacturer's instructions. Fluorescence images were acquired under identical microscope and exposure settings for all groups. Live and dead cells were identified by green Calcein-AM and red EthD-1 fluorescence, respectively.

Cell-associated ZFO MNP accumulation was examined by bright-field microscopy. To start with, the SKOV3 cells were seeded in six-well plates and exposed to 100 or 500 μg/mL ZFO MNPs for 24 h. After treatment, the cells were washed thoroughly with PBS and fixed with 4% paraformaldehyde for 15 min at room temperature. Bright-field images were acquired using an EVOS FL microscope (AMF5000SV, USA). Nanoparticle-associated dark contrast within or adjacent to the cells was interpreted as cell-associated ZFO accumulation.

**2.5. MPS Analysis of Cell-Associated ZFO MNPs**

SKOV3 samples containing $0.1\times10^6$, $1\times10^6$, or $2\times10^6$ cells were prepared to investigate the relationship between cell number and the magnetic signal associated with ZFO exposure. SKOV3 cells were seeded in six-well plates at the specified cell numbers and allowed to attach under the culture conditions described in **Section 2.4**. After attachment, the cells were exposed to ZFO MNPs at a final concentration of 6 mg/mL for 24 h. After incubation, the nanoparticle-containing medium was removed, and the cells were washed with PBS to remove unbound and loosely associated nanoparticles. The cells were then detached, collected, resuspended in FBS, and transferred to glass vials for MPS measurements. This ZFO concentration was substantially higher than the 100 and 500 μg/mL concentrations used for the microscopy experiments in **Section 2.4**. At 6 mg/mL, the dense nanoparticle-associated contrast obscured the cell bodies and pericellular regions in bright-field images. Therefore, these samples were characterized magnetically by MPS instead of microscopy.

Dynamic magnetization responses of ZFO MNPs were measured using a custom-built MPS system reported in our previous work[44–47]. Excitation generation and signal acquisition were controlled using a USB-6289 data-acquisition device (National Instruments, Austin, TX, USA). The excitation waveform was generated at a sampling rate of 1 MHz, and the receive coil voltage was digitized at 625 kHz. The detected signal was amplified using a low-noise voltage preamplifier (SR560; Stanford Research Systems) and band-pass-filtered between 100 Hz and 300 kHz. The preamplifier gain was adjusted for each acquisition to make effective use of the analog-to-digital converter (ADC) input range while avoiding amplifier overload and ADC clipping. The measured waveforms were normalized by the corresponding gain before analysis. ZFO MNPs dispersed in deionized water or immobilized in agar were measured under two different sinusoidal excitation fields: 20 mT at 7.75 kHz and 10 mT at 11.37 kHz. These two excitation frequencies were selected according to the coherent sampling condition $f_0 = kf_s/N$, where $k$ is the integer number of excitation cycles within the acquisition window, $f_s = 625$ kHz is the sampling rate, and $N$ is the number of points. For the 20 mT field, $k = 62$ and $N = 5{,}000$, yielding $f_0 = 7{,}750$ Hz; for the 10 mT field, $k = 149$ and $N = 8{,}192$, yielding $f_0 \approx 11{,}368$ Hz. After subtraction of the background signal recorded with an empty receive coil to further cancel out feed-through, the time-domain voltage signal was converted to the frequency domain, and the higher-order harmonic amplitudes were extracted. For all spectral analyses, a Hann window normalized to unit mean was applied to the time-domain signals before Fourier transformation to reduce spectral leakage while compensating for window-induced amplitude attenuation. To compare spectral shape across samples, normalized harmonic ratios were calculated from the extracted voltage amplitudes as:

$$R_m^n = 20\log_{10}\frac{A_n}{A_m} \quad (1),$$

where $A_n$ and $A_m$ are the amplitudes of the $n^{th}$ and $m^{th}$ voltage harmonics, respectively. Ratios referenced to the fundamental ($m$=1) are expressed in dBc, whereas fifth-to-third and seventh-to-fifth harmonic ratios are expressed in dB. Because biological cells and the suspending medium generated negligible higher-order magnetic harmonics under these excitation conditions, the measured harmonic responses were attributed primarily to ZFO MNPs retained by the cell samples. Harmonic amplitudes were compared among the three cell-number groups to assess relative nanoparticle association. Normalized harmonic ratios and spectral decay were further analyzed to evaluate changes in the ensemble-averaged dynamic magnetic response and physical confinement of the cell-associated ZFO MNPs.

## 3. RESULTS AND DISCUSSIONS

### 3.1. Physicochemical Properties of ZFO MNPs

The morphology, crystal structure, aqueous dispersion, static magnetic properties, and surface chemistry of the synthesized ZFO MNPs were characterized, as summarized in **Figure 1**. The TEM image in **Figure 1a** shows faceted nanoparticles with a predominantly cubic morphology and a characteristic core size of around 30 nm. Some particle-to-particle contact and aggregation are visible in the dried TEM specimen, which may result from magnetic dipole-dipole interactions and particle concentration during grid drying. Nevertheless, the individual particle boundaries remain distinguishable, confirming the formation of nanoscale ZFO cubes. The XRD pattern in **Figure 1b** exhibits the characteristic reflections of a crystalline spinel ferrite phase. No prominent peaks attributable to secondary crystalline phases are observed over the measured 2θ range, supporting the formation of phase-pure ZFO within the detection limit of XRD. The relatively sharp diffraction peaks are consistent with the high crystallinity and comparatively large crystalline dimensions of the hydrothermally synthesized nanoparticles.

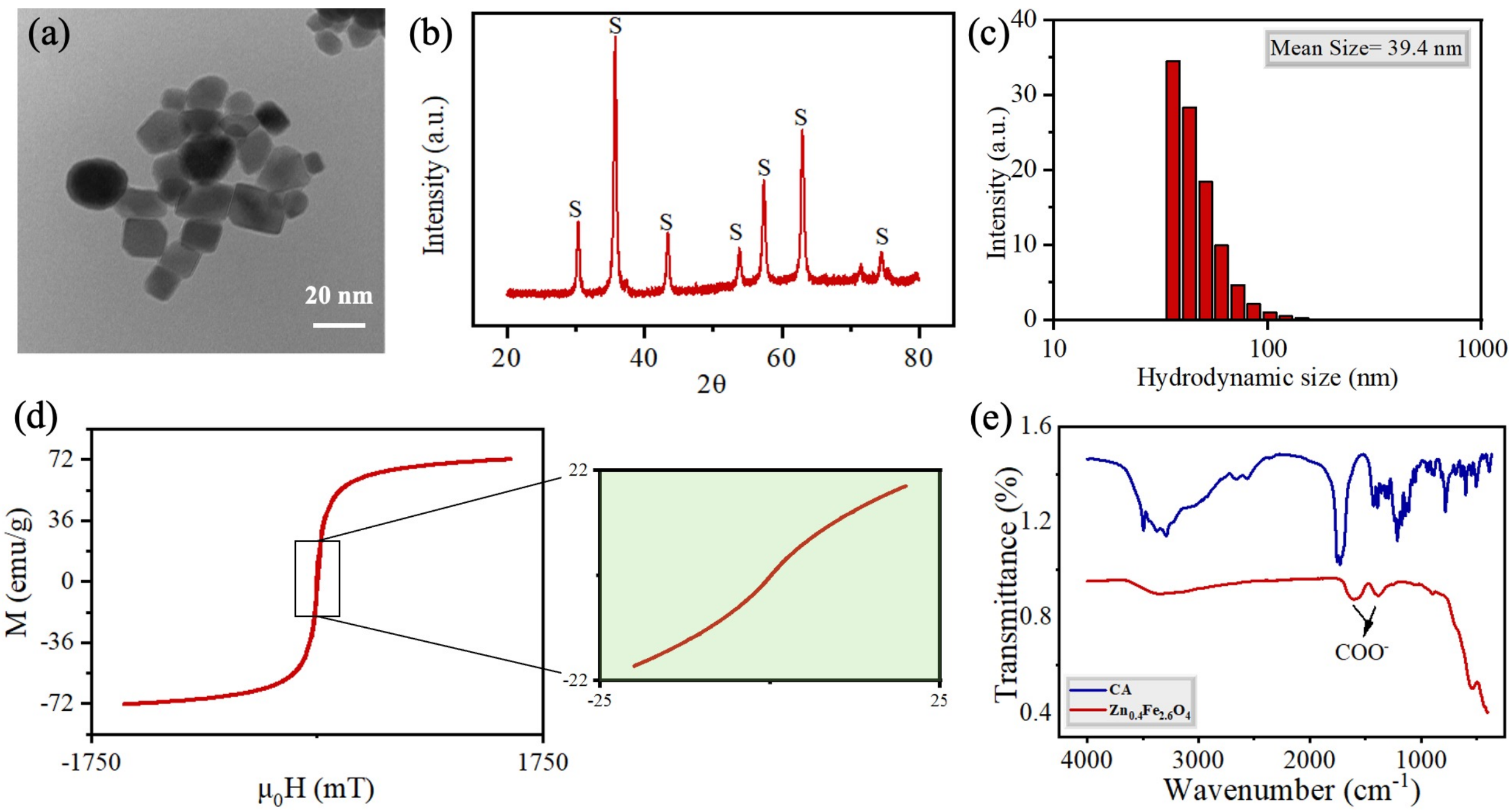

**Figure 1**. Morphological, structural, colloidal, magnetic, and surface-chemical characterization of ZFO MNPs. (a) Representative TEM image of 30 nm cubic ZFO MNPs. Scale bar: 20 nm. (b) Powder XRD pattern confirming the spinel ferrite crystal structure; peaks labeled "S" correspond to characteristic spinel reflections. (c) DLS-derived hydrodynamic size distribution of ZFO MNPs dispersed in DI water, showing a mean hydrodynamic diameter of 39.4 nm. (d) Room-temperature M-H curve measured over an applied magnetic field range of -1.5 to +1.5 T; the inset shows the low-field region. (e) FTIR spectra of citric acid (CA) and citrate-functionalized ZFO MNPs, showing carboxylate-associated absorption bands indicative of citrate coordination on the nanoparticle surface.

The hydrodynamic size distribution of ZFO MNPs dispersed in DI water is presented in **Figure 1c**. DLS analysis yielded a mean hydrodynamic diameter of 39.4 nm, which is moderately larger than the 30 nm magnetic core size observed by TEM. This difference is expected because TEM measures the inorganic core in the dried state, whereas DLS measures the effective particle diameter in suspension, including the surface citrate layer, associated hydration shell, and any small particle clusters. The relatively close agreement between the TEM and DLS dimensions suggests that the citrate-functionalized ZFO MNPs were primarily dispersed as individual particles or small aggregates in DI water.

The room-temperature M-H curve in **Figure 1d** indicates the strong magnetic response of the ZFO MNPs. Magnetization increased rapidly at low fields and approached saturation at higher applied fields, reaching 72 emu/g at 1.5 T. The low-field region shows a narrow hysteresis loop with negligible coercivity and remanence. This magnetically soft behavior indicates that the ZFO MNPs can readily reverse their magnetization in response to an external field while retaining little residual magnetization after field removal. The combination of high magnetization and low coercivity is favorable for generating strong, rapidly changing magnetic responses under the excitation fields in MPS.

FTIR spectroscopy was used to examine the surface chemistry of the ZFO MNPs. As shown in **Figure 1e**, the citrate-functionalized nanoparticles exhibit carboxylate-related absorption features in the approximately 1400-1600 $cm^{-1}$ region. The presence of these $COO^-$ bands supports the association of citrate groups with the nanoparticle surfaces, likely through coordination between carboxylate groups and surface metal ions. Absorption at lower wavenumbers is attributed primarily to metal-oxygen vibrations of the ferrite lattice. Citrate functionalization introduces hydrophilic, negatively charged surface groups that improve aqueous dispersibility and provide sites for subsequent bioconjugation or surface modification.

### 3.2. Cytotoxicity and Cellular Uptake of MNPs

The short-term cytocompatibility of the ZFO MNPs was evaluated in SKOV3 ovarian cancer cells using a fluorescence-based Live/Dead assay after 24 h of exposure. As shown in **Figure 2a&b**, cells treated with 100 and 500 μg/mL ZFO MNPs exhibited predominantly green Calcein-AM fluorescence, indicating that most cells remained metabolically active with intact cell membranes. Only minimal red EthD-1 fluorescence, which indicates membrane-compromised dead cells, was observed at either concentration. Moreover, increasing the ZFO concentration from 100 to 500 μg/mL did not produce an obvious increase in dead-cell staining. These qualitative observations suggest that the citrate-functionalized ZFO MNPs did not induce acute cytotoxicity in SKOV3 cells under the tested exposure conditions.

Cell-associated nanoparticle accumulation was subsequently examined by bright-field microscopy. Following 24 h of incubation, dark particulate contrast was visible within the cell-containing regions at both 100 and 500 μg/mL, as shown in **Figure 2c&d**, respectively. The observed contrast appeared to be localized near the cell bodies and pericellular regions, demonstrating that the ZFO MNPs remained associated with the cells after removal of the nanoparticle-containing medium and PBS washing. More extensive dark contrast was observed at 500 μg/mL (**Figure 2d**) than at 100 μg/mL (**Figure 2c**), indicating greater cell-associated nanoparticle accumulation at the higher exposure concentration. This concentration-dependent behavior arises from increased nanoparticle contact with the cell membrane, surface adsorption, and/or cellular internalization.

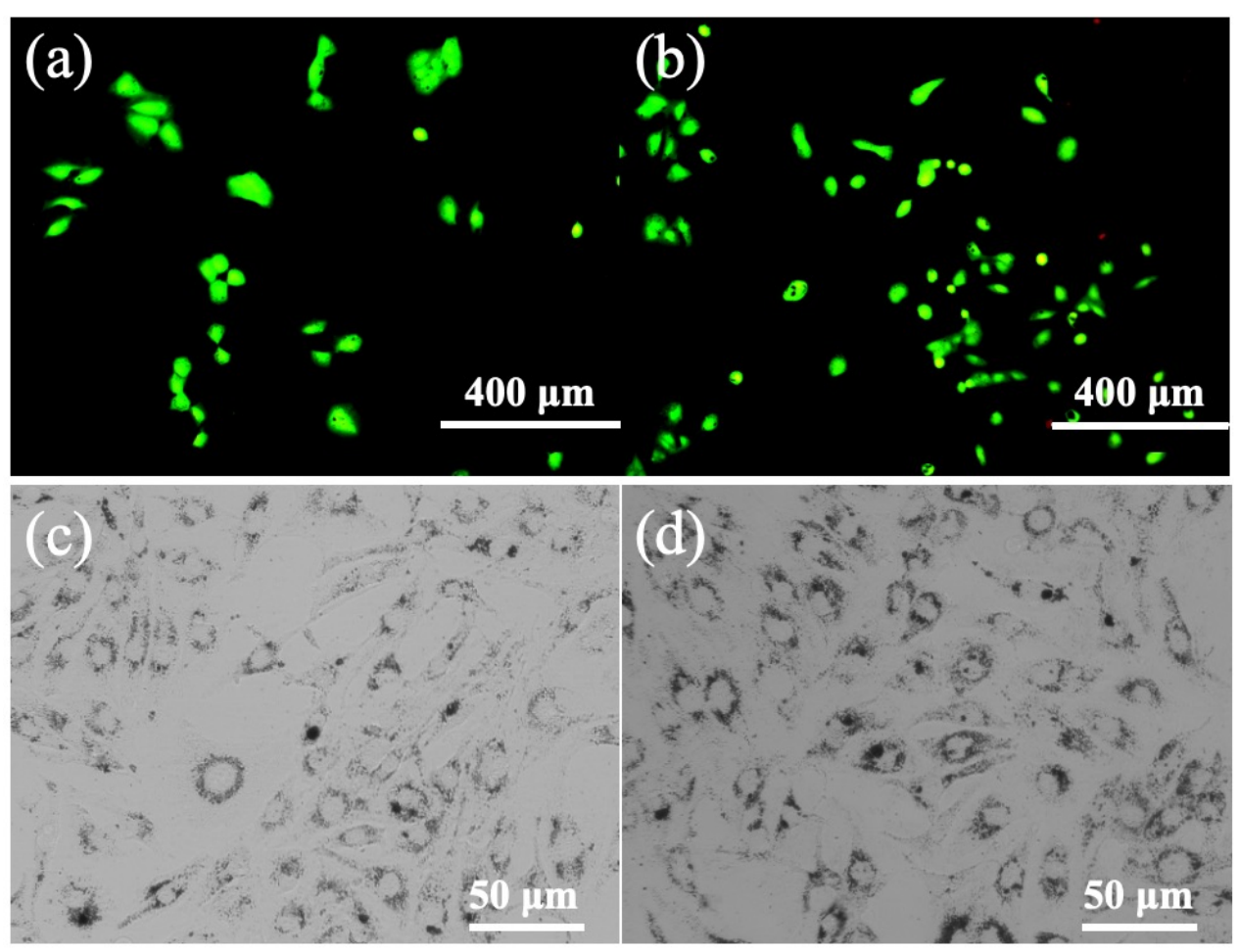

**Figure 2.** Cytocompatibility and cell-associated accumulation of ZFO MNPs in SKOV3 ovarian cancer cells. Fluorescence-based Live/Dead assay after 24 h of incubation with ZFO MNPs at (a) 100 μg/mL and (b) 500 μg/mL. Green Calcein-AM fluorescence indicates viable cells, whereas red EthD-1 fluorescence indicates membrane-compromised cells. Scale bars: 400 μm. Representative bright-field images of SKOV3 cells after 24 h of incubation with ZFO MNPs at (c) 100 μg/mL and (d) 500 μg/mL. Dark contrast indicates cell-associated nanoparticle accumulation. Scale bars: 50 μm.

It should be noted that bright-field microscopy alone cannot distinguish nanoparticles internalized by the cells from those strongly adsorbed to the cell surface. Therefore, the dark contrast in **Figure 2c&d** is interpreted as cell-associated ZFO accumulation. Nevertheless, the persistence of nanoparticle-associated contrast after washing, together with the predominantly viable cell population observed in the Live/Dead assay, indicates that SKOV3 cells can tolerate and retain ZFO MNPs over a 24 h exposure period. These results support the subsequent use of MPS to detect and compare the magnetic responses of ZFO-labeled SKOV3 cell samples.

### 3.3. MPS Characterization on Cell-Associated ZFOs

#### 3.3.1. Drive Field Effect on Higher Harmonic Retention

To select an appropriate excitation condition for the cellular measurements, the dynamic responses of ZFO MNPs were compared under two drive field conditions: 20 mT at 7.75 kHz and 10 mT at 11.37 kHz. ZFO MNPs dispersed in DI water and immobilized in agar were used as reference states with different degrees of rotational freedom. The aqueous suspension represents a relatively unrestricted, low-viscosity environment in which Brownian rotation can contribute to the MPS response, whereas the agar matrix restricts whole-particle rotation and suppresses the Brownian contribution. Cell-associated ZFO MNPs are expected to experience partial physical confinement through membrane association, aggregation, or confinement within intracellular compartments. Accordingly, the water and agar samples were used as phenomenological reference states for interpreting environmentally induced changes in the MPS spectra of ZFO-labeled cells.

**Figure 3** presents the MPS responses measured at 7.75 kHz and 20 mT. The time-domain signal of ZFO MNPs dispersed in DI water exhibits a strongly distorted, sharply peaked waveform, indicating a pronounced nonlinear magnetic response under the 20 mT excitation field (**Figure 3a**). The corresponding frequency-domain spectrum contains detectable odd harmonics through at least the 13$^{th}$ order, with the amplitude decreasing progressively as harmonic order increases (**Figure 3b**). This broad harmonic content indicates that the 20 mT field drives the ZFO magnetization sufficiently far into the nonlinear region of its magnetization curve.

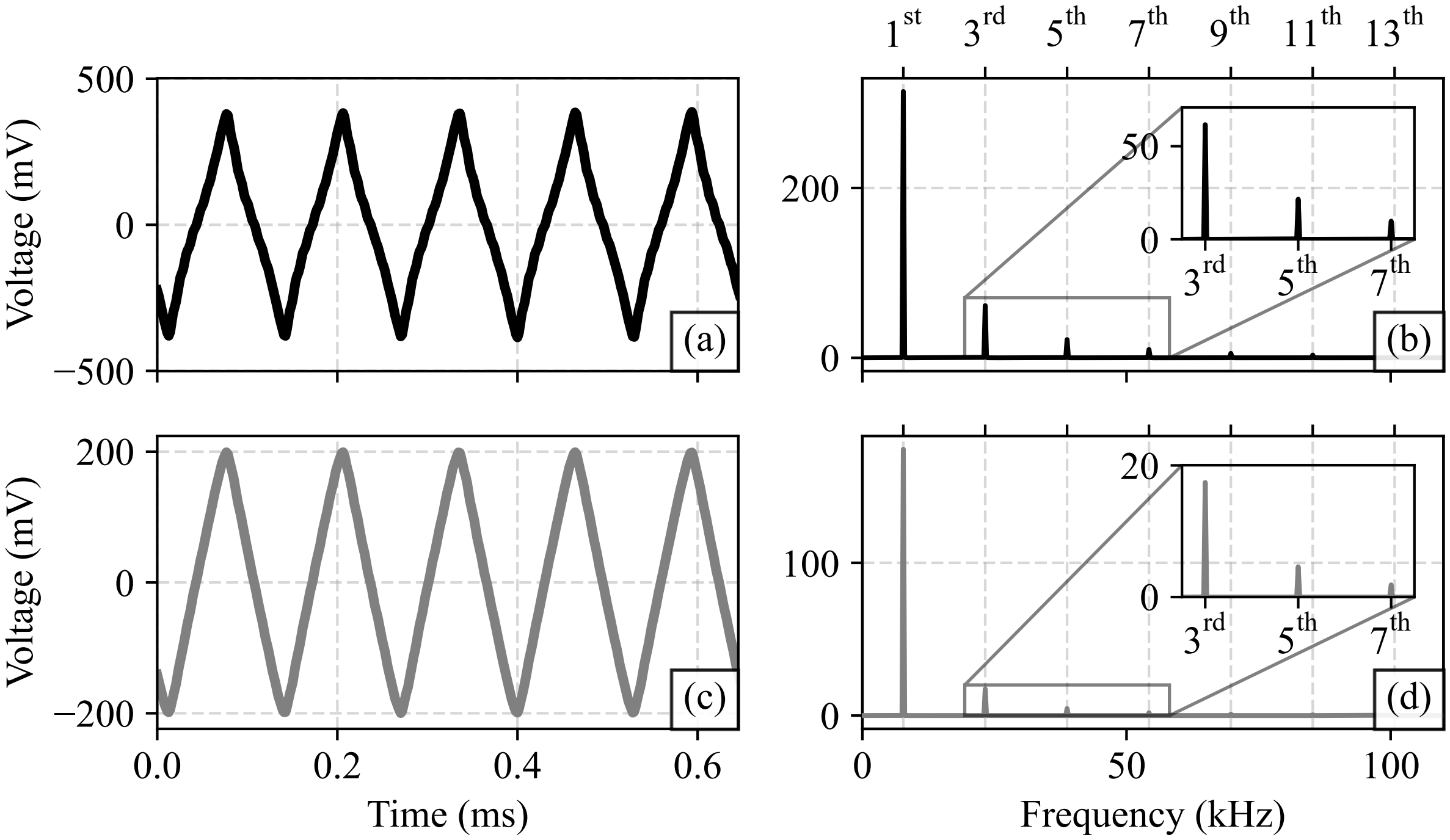

**Figure 3**. MPS responses of 6 mg/mL ZFO MNPs under a drive field of 20 mT at 7.75 kHz. Time-domain induced-voltage waveforms for nanoparticles (a) dispersed in DI water and (c) immobilized in agar, with corresponding frequency-domain spectra shown in (b) and (d), respectively. Insets enlarge the third-, fifth-, and seventh-harmonic components.

Immobilization in agar substantially modifies the dynamic response. The time-domain waveform remains nonlinear but has a lower amplitude and less pronounced distortion than that measured in water (**Figure 3c**). Correspondingly, the harmonic amplitudes are reduced throughout the measured spectrum (**Figure 3d**). These changes are consistent with restriction of Brownian rotation by the agar matrix, leaving the response dominated by Néel relaxation and any limited residual particle mobility within the gel.

**Figure 4** shows the corresponding measurements obtained at 11.37 kHz and 10 mT. Although increasing the excitation frequency can increase the induced receive coil voltage for an otherwise equivalent magnetization response, the effects of frequency and field amplitude cannot be separated in this comparison because both parameters were changed simultaneously. Under the 11.37 kHz and 10 mT condition, the time-domain signals from both the aqueous and agar samples (**Figure 4a,c**) are more nearly sinusoidal than those measured at 7.75 kHz and 20 mT. Their frequency-domain spectra are dominated by the fundamental component, while only relatively weak third-, fifth-, and seventh-order harmonics are retained (**Figure 4b,d**).

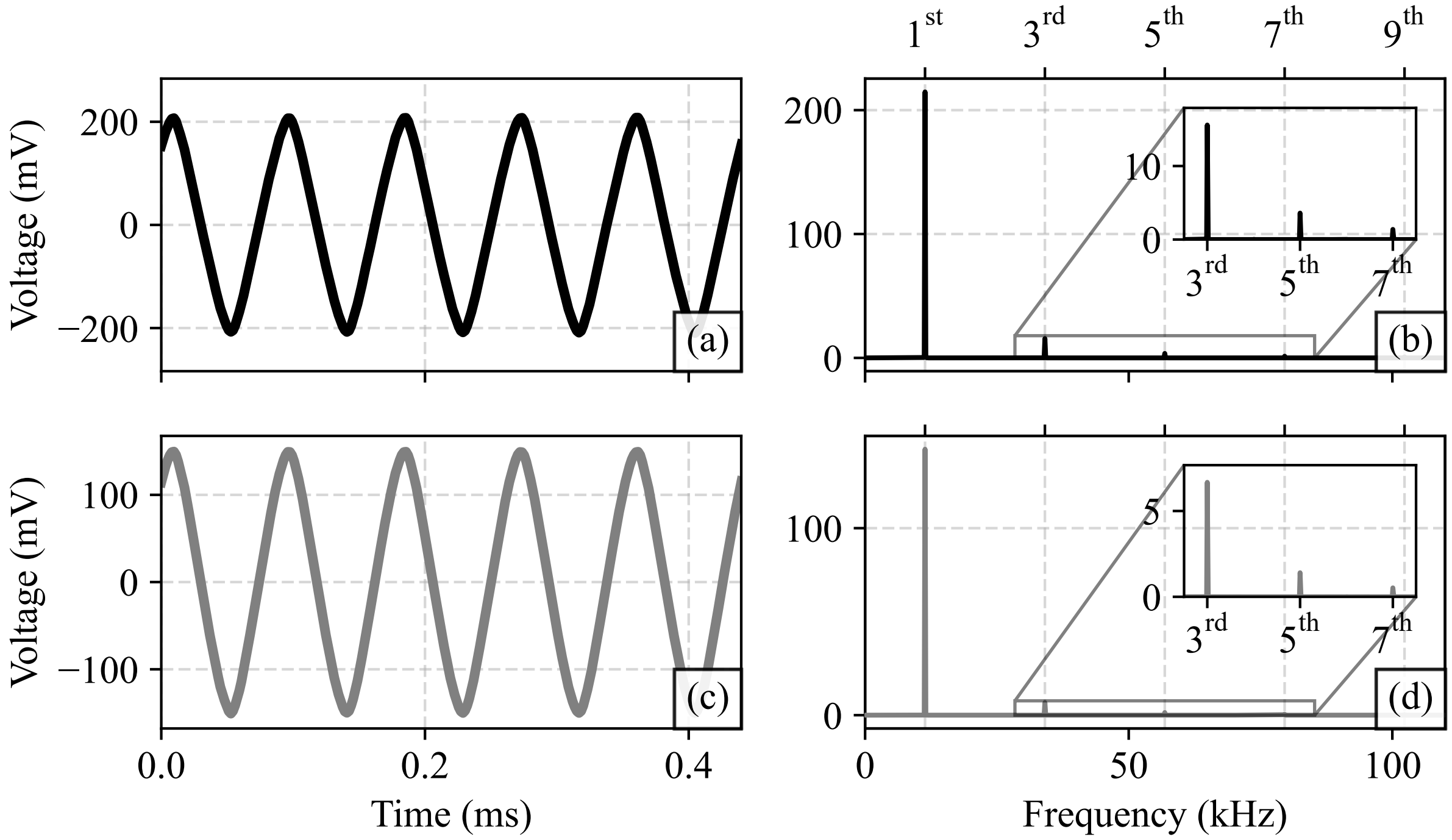

**Figure 4**. MPS responses of 6 mg/mL ZFO MNPs under a drive field of 10 mT at 11.37 kHz. Time-domain induced-voltage waveforms for nanoparticles (a) dispersed in DI water and (c) immobilized in agar, with corresponding frequency-domain spectra shown in (b) and (d), respectively. Insets enlarge the third-, fifth-, and seventh-harmonic components.

As shown in **Figure 5a**, under the 20 mT at 7.75 kHz excitation field, the normalized harmonic ratios to the fundamental frequency component, $R_1^3$, $R_1^5$, and $R_1^7$, are consistently lower for agar-embedded particles than for particles dispersed in DI water. Moreover, the ratios decrease more rapidly with higher harmonic order after agar immobilization. Thus, physical confinement not only reduces the absolute MPS harmonic signals but also suppresses the higher-order components relative to the fundamental. These normalized ratios provide spectral metrics for subsequent comparison with ZFO-labeled cell samples. Under the excitation field of 10 mT at 11.37 kHz, the normalized harmonic ratios (**Figure 5b**) are also substantially lower than those obtained under the 20 mT condition, confirming more rapid attenuation of the higher-order harmonics, and the water-agar gap remains small and nearly constant across $R_1^3$ to $R_1^7$.

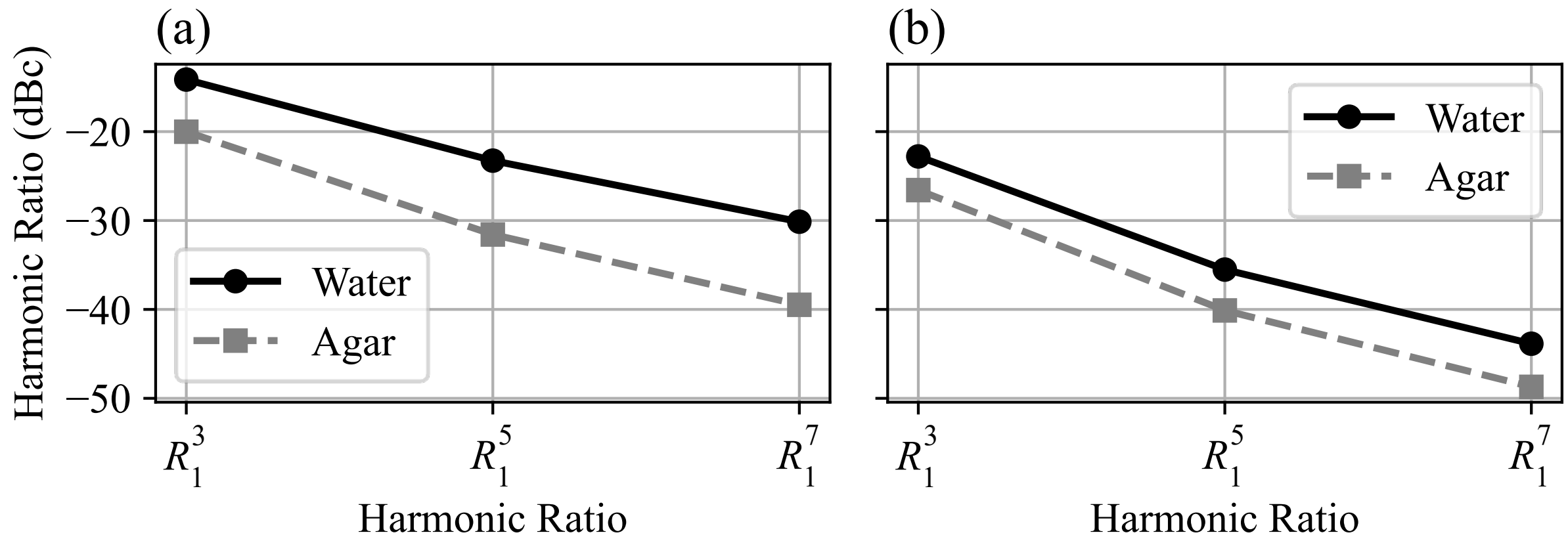

**Figure 5**. Normalized harmonic ratios of ZFO MNPs dispersed in DI water or immobilized in agar under excitation fields of (a) 20 mT at 7.75 kHz and (b) 10 mT at 11.37 kHz.

In summary, the 7.75 kHz, 20 mT excitation field retains a broader range of detectable higher harmonics from ZFO MNPs and produces a clearly larger and order-dependent spectral separation between freely dispersed and agar-immobilized states. This condition was therefore selected for the subsequent MPS measurements of ZFO-labeled SKOV3 cells, and the DI water and agar harmonic ratios were used as reference profiles representing relatively unrestricted and strongly confined particle environments, respectively.

#### 3.3.2. MPS Characterization of Cell-Associated ZFO MNPs

The dynamic magnetic responses of ZFO-labeled SKOV3 samples containing 2, 1, or 0.1 million cells were recorded on MPS under the selected excitation field condition of 7.75 kHz and 20 mT. As described in **Section 2.5**, free and loosely associated ZFO MNPs were removed before MPS measurements. Because unlabeled cells and the suspending medium generated negligible higher-order magnetic harmonics under these conditions, the detected nonlinear signals were attributed primarily to ZFO MNPs retained by the cell samples. **Figure 6** presents the time- and frequency-domain MPS signals from the three ZFO-labeled SKOV3 cell samples. The sample containing $2\times10^6$ cells produced a strongly distorted periodic waveform with a peak voltage of 6.1 mV (**Figure 6a**). Its corresponding frequency-domain spectrum contained a prominent fundamental component and clearly detectable higher odd harmonics, including the third, fifth, and seventh orders (**Figure 6b**). The presence of these higher harmonics confirms that a measurable quantity of magnetically responsive ZFO MNPs remained associated with the cells after the post-incubation washing procedure.

The sample containing $1\times10^6$ cells also exhibited a nonlinear time-domain waveform, although its voltage amplitude was lower than that of the $2\times10^6$ cell sample (**Figure 6c**). Correspondingly, the fundamental and higher odd-harmonic amplitudes were reduced (**Figure 6d**). Nevertheless, the third-, fifth-, and seventh-order components remained distinguishable from the spectral background, demonstrating that MPS could detect cell-associated ZFO MNPs in the $1\times10^6$-cell sample. A further reduction in MPS signal was observed for the sample containing $0.1\times10^6$ cells.

Its time-domain response had a substantially lower amplitude and a visibly higher relative noise level than those of the two larger cell samples (**Figure 6e**). The frequency-domain spectrum likewise showed reduced harmonic amplitudes, although ZFO-associated components remained detectable at the lower odd harmonics (**Figure 6f**). The reduced signal is consistent with the smaller number of cells available to retain ZFO MNPs following incubation and washing.

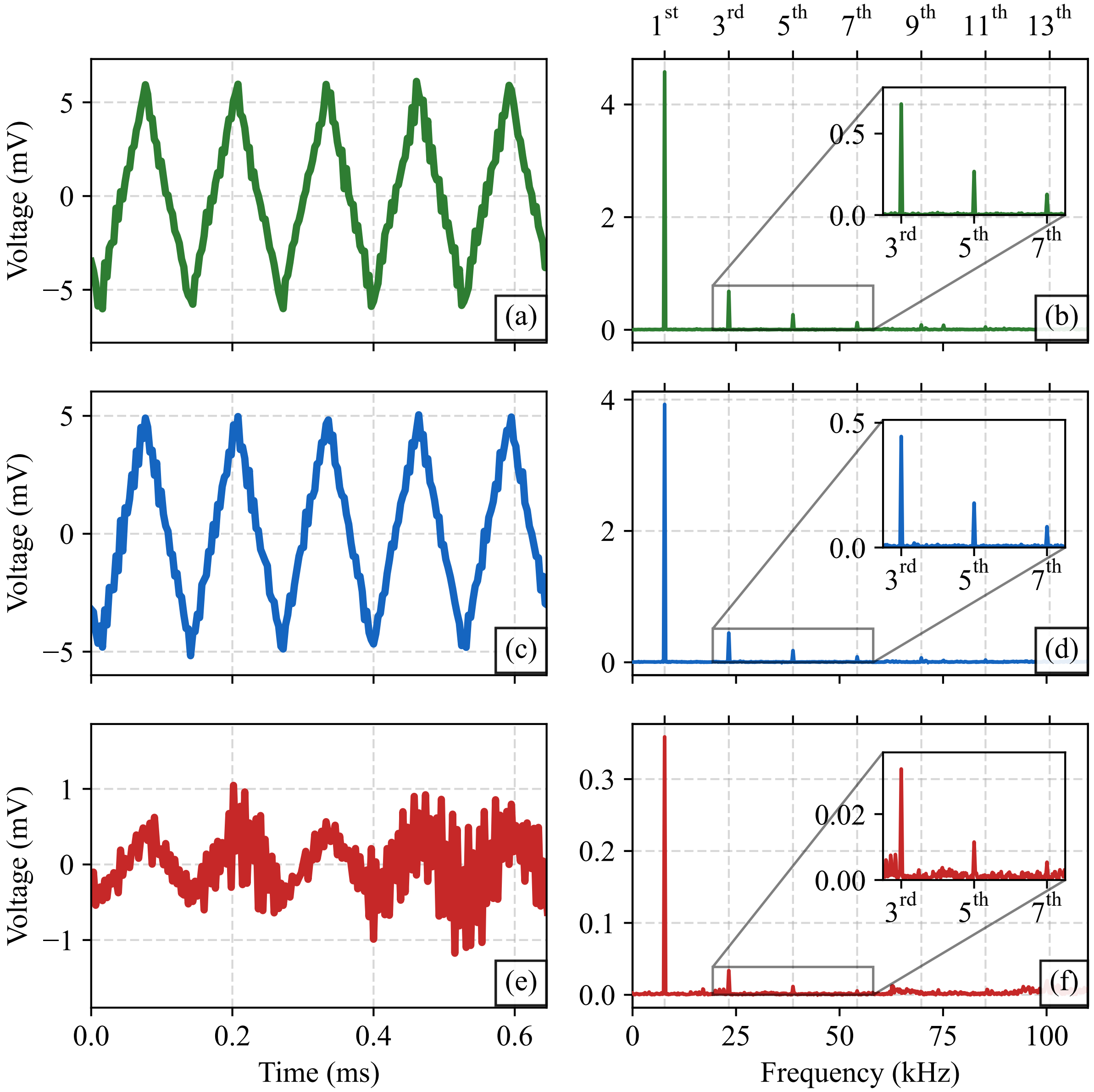

**Figure 6**. MPS responses of ZFO-labeled SKOV3 cell samples measured under a drive field of 7.75 kHz and 20 mT. Time-domain induced-voltage signals from samples containing (a) $2\times10^6$, (c) $1\times10^6$, and (e) $0.1\times10^6$ SKOV3 cells after incubation with ZFO MNPs and removal of unbound nanoparticles. The corresponding frequency-domain spectra are shown in (b), (d), and (f), respectively; the insets enlarge the detectable higher odd-harmonic components.

Overall, the results in **Figure 6** demonstrate that MPS can detect ZFO MNPs retained by SKOV3 cells across the tested cell-number range. The dependence of the absolute harmonic amplitudes and normalized spectral profiles on cell number is quantitatively compared in **Figure 7**. As summarized in **Figure 7a**, the third- and fifth-harmonic amplitudes increased monotonically with cell number, with the $2\times10^6$-cell sample producing the strongest response, followed by the $1\times10^6$- and $0.1\times10^6$-cell samples. It should be noted that the MPS signal reflects the total amount of ZFO MNPs recovered with each cell population rather than cell number directly; therefore, differences in nanoparticle loading per cell, cell recovery during sample preparation, and strongly surface-associated nanoparticles may also influence the measured amplitudes. For all three samples, harmonic amplitude decreased with increasing harmonic order, consistent with the progressive spectral attenuation of the nonlinear magnetization response. When normalized to the fundamental component, $R_1^3$, $R_1^5$, and $R_1^7$ exhibited similar decreasing trends across the three samples (**Figure 7b**). The fifth-to-third and seventh-to-fifth harmonic ratios provide an additional comparison of spectral shape (**Figure 7c**). The $2\times10^6$- and $1\times10^6$-cell samples showed broadly comparable ratios, suggesting similar ensemble-averaged dynamic magnetic environments. The $0.1\times10^6$-cell sample exhibited a larger deviation, particularly in the fifth-to-third harmonic ratio. Because this sample generated the weakest absolute response, the deviation is likely influenced by its lower signal-to-noise ratio and should not be interpreted as definitive evidence of a different nanoparticle relaxation state.

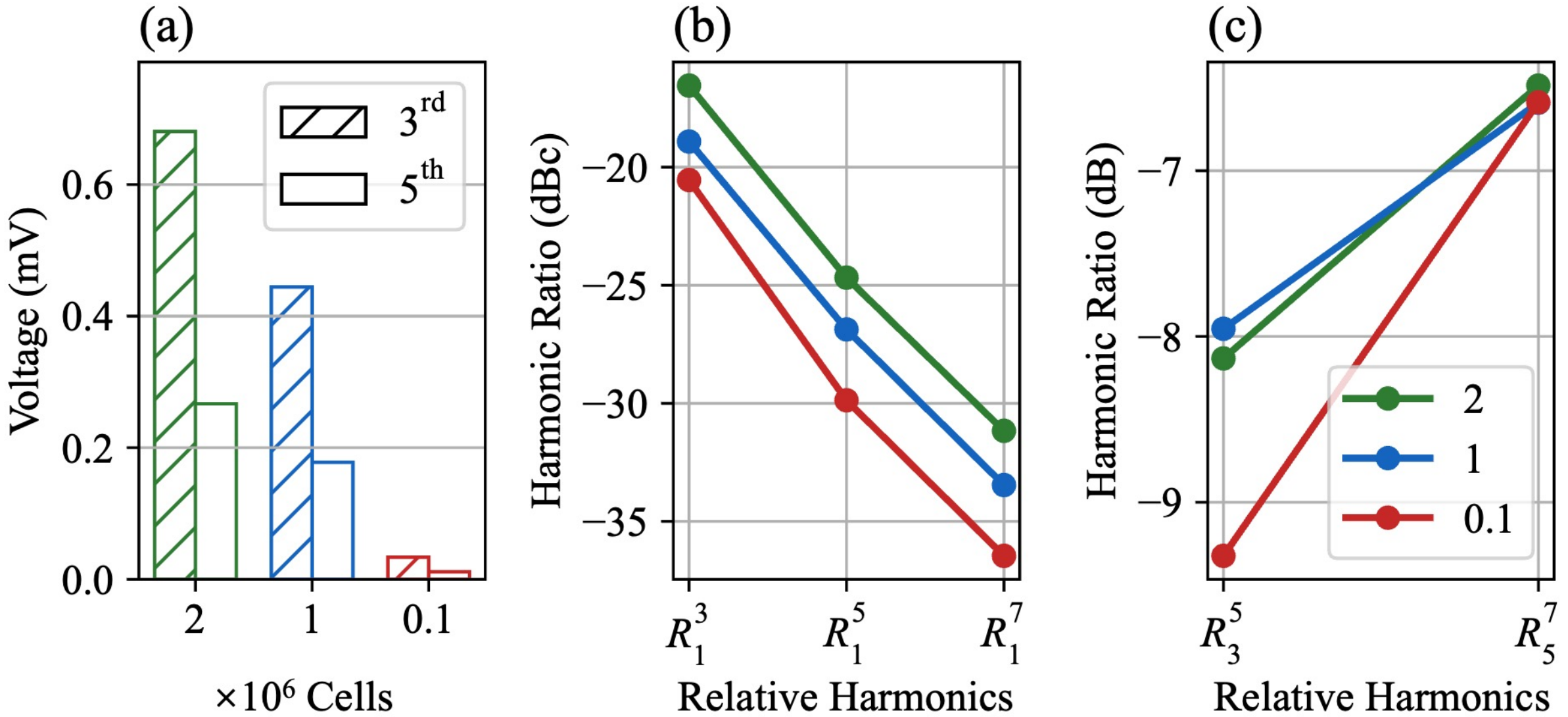

**Figure 7**. (a) Summary of the third- and fifth-harmonic amplitudes recorded from ZFO-labeled SKOV3 cell samples containing $2\times10^6$, $1\times10^6$, and $0.1\times10^6$ SKOV3 cells. (b) Third-, fifth-, and seventh-to-fundamental harmonic ratios for ZFO-labeled SKOV3 cells. Fifth-to-third and seventh-to-fifth harmonic ratios for ZFO-labeled SKOV3 cells.

**Figure 8** shows the relationship between cell number and the third-, fifth-, and seventh-harmonic amplitudes measured from ZFO-labeled SKOV3 samples. Over the tested range of $0.1\times10^6$ to $2\times10^6$ cells, all three harmonic amplitudes increased systematically with cell number.

Power-law fitting using $V_n = A_n X^{B_n}$, where $X$ is the cell number expressed in units of $10^6$ cells, yielded exponents of 1.033, 1.085, and 1.088 for the third, fifth, and seventh harmonics, respectively, with corresponding $R^2$ values of 99.17%, 98.92%, and 98.99%. Because an exponent of 1 represents direct proportionality, these results indicate an approximately linear increase in harmonic amplitude with cell number, particularly for the third harmonic. The slightly larger exponents obtained for the fifth and seventh harmonics should not be interpreted as definitive superlinear behavior because only three cell-number levels were evaluated and the higher-order components had lower signal-to-noise ratios. Moreover, each culture was exposed to the same total ZFO mass instead of an equivalent nanoparticle dose per cell. Therefore, variations in labeling efficiency, cell recovery, and average nanoparticle loading may contribute to the measured relationship.

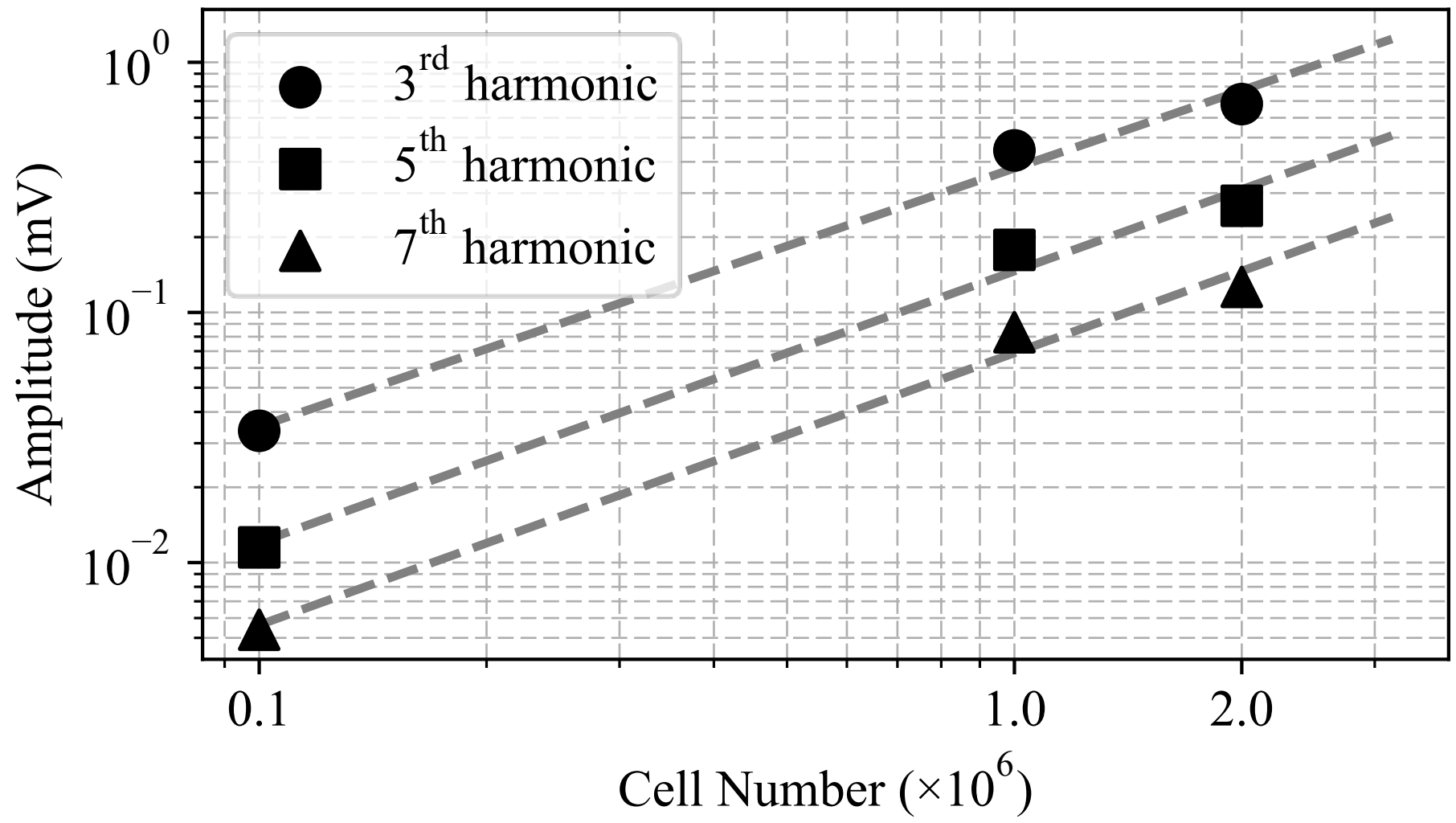

**Figure 8**. Relationship between MPS harmonic amplitude and the number of ZFO-labeled SKOV3 cells. Log-log plots of the third-, fifth-, and seventh-harmonic amplitudes as functions of cell number, $X$, expressed in units of $\times 10^6$ cells. Dashed lines show power-law fits according to $V_n = A_n X^{B_n}$. The fitted equations were $V_3 = 0.377X^{1.033}$ ($R^2 = 0.992$), $V_5 = 0.146X^{1.085}$ ($R^2 = 0.989$), and $V_7 = 0.069X^{1.088}$ ($R^2 = 0.990$), where the harmonic amplitudes are expressed in mV.

Unlike absolute harmonic amplitudes, which scale with the amount of magnetically detectable ZFO MNPs, normalized harmonic ratios are expected to be comparatively insensitive to nanoparticle quantity when the measurement system operates within its linear range, and the relevant harmonics remain above the noise floor. Instead, these ratios describe the relative shape of the harmonic spectrum and provide information about the ensemble-averaged dynamic magnetization and relaxation behavior of the nanoparticles. The ratios can therefore be used to compare ZFO MNPs across samples containing different total nanoparticle amounts. However, they may also be affected by measurement noise, nanoparticle aggregation, magnetic interactions, and heterogeneity in the local particle environment. **Figure 9** overlays the harmonic ratios obtained from the three ZFO-labeled SKOV3 cell samples with the range defined by ZFO MNPs dispersed in DI water and immobilized in agar. The aqueous sample represents a relatively unrestricted, low-viscosity state in which Brownian rotation can contribute substantially to the dynamic magnetic response. In contrast, the agar matrix strongly restricts whole-particle rotation and suppresses the Brownian contribution. These two conditions were therefore used as phenomenological reference states rather than exact models of the intracellular environment.

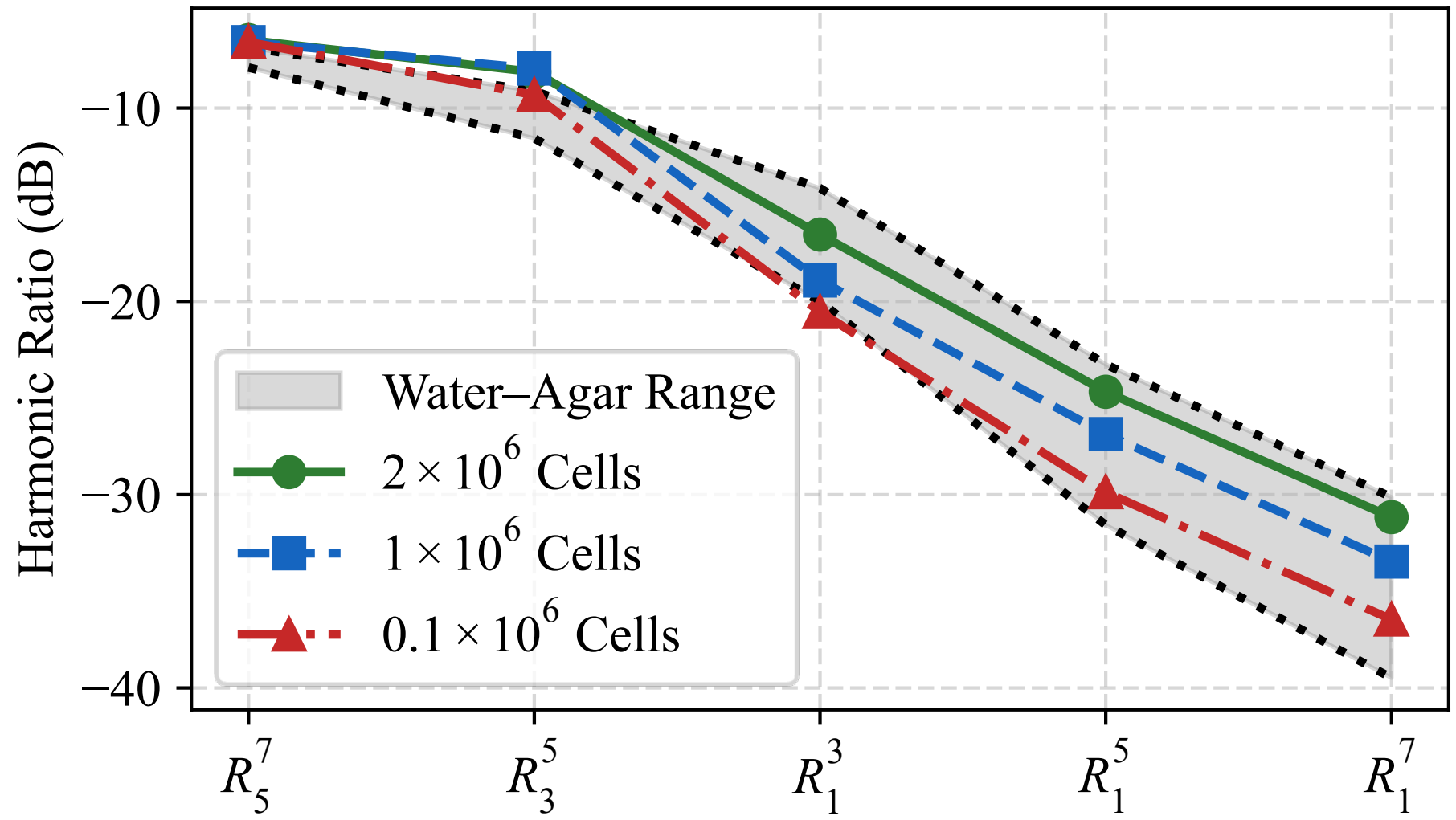

**Figure 9**. Comparison of harmonic ratios from ZFO-labeled SKOV3 cells and ZFO MNPs in DI water and agar reference environments. The gray region denotes the range bounded by ZFO MNPs dispersed in DI water and immobilized in agar, representing relatively unrestricted and strongly restricted Brownian rotation, respectively. Harmonic ratios from samples containing $2\times10^6$, $1\times10^6$, and $0.1\times10^6$ ZFO-labeled SKOV3 cells are overlaid for comparison.

Most harmonic ratios measured from the $1\times10^6$- and $2\times10^6$-cell samples fall within the water-agar reference range, suggesting that the cell-associated ZFO MNPs exhibit an ensemble-averaged dynamic state intermediate between freely dispersed and strongly immobilized particles. This result qualitatively suggests that cell-associated ZFO MNPs experience partial restriction of Brownian rotation, with an ensemble-averaged physical state intermediate between freely dispersed particles in water and strongly immobilized particles in agar.

## 4. CONCLUSION

In this study, citrate-functionalized ZFO MNPs with an ~30 nm cubic morphology were synthesized and evaluated for MPS-based detection of cell-associated nanoparticles. The ZFO MNPs exhibited a crystalline spinel ferrite structure, a mean hydrodynamic diameter of 39.4 nm in DI water, strong room-temperature magnetization, low coercivity, and citrate-associated surface functional groups. Live/Dead imaging further indicated good short-term cytocompatibility in SKOV3 ovarian cancer cells after 24 h of exposure at concentrations up to 500 µg/mL. Bright-field microscopy showed concentration-dependent cell-associated nanoparticle accumulation, although this method could not distinguish internalized particles from those strongly adsorbed to the cell surface.

Comparison of two MPS excitation conditions demonstrated that a drive field of 7.75 kHz and 20 mT produced stronger nonlinear responses and better retention of higher-order harmonics than 11.37 kHz and 10 mT. Under the selected condition, cell-associated ZFO MNPs were detected in samples containing $0.1\times10^6$ to $2\times10^6$ SKOV3 cells. The third-, fifth-, and seventh-harmonic amplitudes increased approximately proportionally with cell number, with power-law exponents of 1.03-1.09. These relationships describe the total magnetically detectable ZFO retained by each cell sample rather than cell number directly. Normalized harmonic ratios provided complementary information about the ensemble-averaged relaxation behavior of the cell-associated nanoparticles. Most ratios from the $1\times10^6$- and $2\times10^6$-cell samples fell within the range defined by ZFO MNPs dispersed in water and immobilized in agar. This finding qualitatively suggests that cell-associated ZFO MNPs experience partial restriction of Brownian rotation, producing a dynamic state intermediate between relatively unrestricted aqueous dispersion and strong agar confinement. Deviations observed for the $0.1\times10^6$-cell sample were attributed primarily to the lower signal-to-noise ratio of its higher-order harmonics.

In summary, our findings demonstrate that MPS can detect cell-associated ZFO MNPs while simultaneously providing spectral information related to their physical confinement. Future studies incorporating biological replicates, additional cell-number levels, quantitative nanoparticle-uptake measurements, and complementary imaging of intracellular localization will be needed to establish a validated calibration between MPS signal and labeled-cell number and to clarify the intracellular relaxation behavior of ZFO MNPs.

## ASSOCIATED CONTENT

### Author Contributions

K.W. conceptualized the study, acquired funding, and supervised the project. B.R. synthesized and characterized nanoparticles. H.W. performed the MPS characterizations. M.S. and C.X. conducted the cytotoxicity and cellular uptake studies. H.W. led the manuscript drafting. R.H. and K.W. led the data interpretation. K.W. contributed to manuscript writing, revision, and final proofreading.

### Funding

Research reported in this publication was supported by the National Institute Of Biomedical Imaging And Bioengineering of the National Institutes of Health under Award Number R03EB036435. The content is solely the responsibility of the authors and does not necessarily represent the official views of the National Institutes of Health. This material is based upon work supported by the U.S. National Science Foundation under award No. 2630047. Any opinions, findings and conclusions or recommendations expressed in this material are those of the author(s) and do not necessarily reflect the views of the U.S. National Science Foundation.

### Notes

The authors declare no competing financial interests.

### Data Availability Statement

Data supporting this study are openly available on Zenodo at:
https://doi.org/10.5281/zenodo.22717769

**TOC Graphic:**

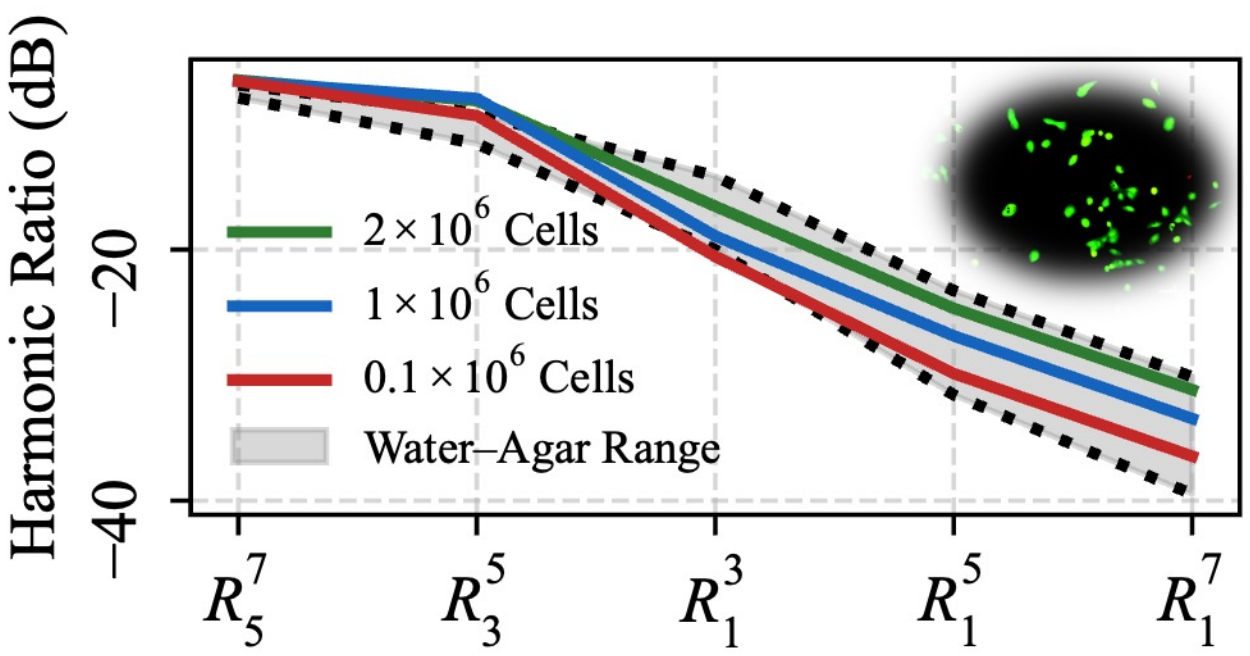

Harmonic Ratio (dB)
−20
−40
2 × 10^6 Cells
1 × 10^6 Cells
0.1 × 10^6 Cells
Water–Agar Range
$R_5^7$
$R_3^5$
$R_1^3$
$R_1^5$
$R_1^7$